\let\TeXyear\year

\documentclass[doublecol]{epl2} 

\let\year\TeXyear

\usepackage{amssymb}
\usepackage{graphicx} 
\usepackage{dcolumn}        
\usepackage{bm}            
\usepackage[normalem]{ulem}
\usepackage{amsmath}
\usepackage{subfigure} 
\usepackage{soul,xcolor}
\usepackage{microtype}
\usepackage{orcidlink}
\usepackage{multirow}
\usepackage{flushend}

\title{A simple light-trapping device from a hyperbolic metamaterial on a catenoid}
\shorttitle{
A simple light-trapping device from a hyperbolic metamaterial on a catenoid} 

\author{Frankbelson dos Santos Azevedo\orcidlink{0000-0002-4009-0720} \inst{1} \and Jos\'{e} Di\^{e}go M. de Lima\orcidlink{0000-0002-4296-6001} \inst{1,2} \and Ant\^{o}nio de P\'{a}dua Santos\orcidlink{0000-0003-1262-0875} \inst{1}  \and Tiago A. E. Ferreira\orcidlink{0000-0002-2131-9825} \inst{3} \and Fernando Moraes\orcidlink{0000-0001-7045-054X} \inst{1}} 

\shortauthor{Frankbelson dos Santos Azevedo \etal}

\institute{               
  \inst{1} Departamento de F\'{\i}sica, Universidade Federal Rural de Pernambuco, 52171-900, Recife, PE, Brazil \\
  \inst{2} Departamento de F\'{\i}sica, Universidade Federal de Pernambuco, 50670-901, Recife, PE, Brazil \\
  \inst{3} Departamento de Estat\'{\i}stica e Informática, Universidade Federal Rural de Pernambuco, 52171-900, Recife, PE, Brazil
}


\abstract{By using both ray and wave optics, we show that a simple device which consists on a film of hyperbolic metamaterial on the surface of a catenoid can be used to trap light. From the study of the trajectories, we observe a tendency for the light rays to wrap, and eventually be trapped, around the neck of the device. The wave equation appears to have an effective attractive potential, and their solutions confirm the bound states suggested by the trajectories. The relevant equations are solved numerically using neural networks.}

\begin{document}

\maketitle

Metamaterials are a class of optical materials that may present a negative ratio between the refractive index components \cite{smith2004metamaterials}. Hence, they may provide many applications for technological devices and new discoveries in the science of transformation optics \cite{chen2010transformation}. From metamaterial devices, we can get analogies with cosmological systems as a way to verify and discuss ideas of cosmology in the laboratory. For example, with metamaterials, it is possible to mimic curved spacetime \cite{genov2009mimicking}, to observe gravitational lens by trapping light \cite{sheng2013trapping}, as well as to simulate black holes \cite{fernandez2016anisotropic} and spinning cosmic string spacetime \cite{mackay2010towards}. Also, they may be used to mimic a discontinuous change of metric signature \cite{figueiredo2016modeling}.

An important class of metamaterials, called ``hyperbolic'' \cite{Guo},  can be found in nature \cite{narimanov2015metamaterials} but are mostly artificial. These materials can be realized through layered metal–dielectric structures \cite{poddubny2013hyperbolic} or, alternatively, from nematic liquid crystals  with metallic nanorods mixed in \cite{xiang2012liquid}. Among the applications of such materials, we mention the design of hyperlenses \cite{jacob2006optical} and analogies with cosmological systems \cite{figueiredo2017cosmology}. In a study made by two of us and coworkers, a hyperbolic liquid crystal metamaterial  with molecules circularly arranged in a cylinder, shows  optical concentrator behavior, for light is focused on its axis while  propagating along the  device \cite{azevedo2018optical}. In addition, wormhole representations based on metamaterials are well-known in the literature; Maslovski et al. \cite{maslovski2018superabsorbing} showed that superabsorbent metamaterial wormholes are conceivable with meshes of loaded transmission lines. Moreover, electromagnetic wormholes from metamaterials objects behave as virtual magnetic monopoles \cite{greenleaf2007electromagnetic}: electromagnetic waves propagate between two points in space through an invisible tunnel. This was later experimentally demonstrated by using magnetic metamaterials and metasurfaces \cite{prat2015magnetic}.

In a recent work \cite{azevedo2021optical}, we showed that an oriented thin nematic liquid crystal film on a catenoid has an optical metric that corresponds to the geometry of a two-dimensional section of a conical wormhole spacetime. In that article, it was suggested the existence of zero angular momentum bound states, which would be soon verified by Atanasov et al. \cite{10.1088/1402-4896/ac1991}, that showed the possibility of having wormholes as waveguides for quantum particles with zero angular momentum, having an optical wormhole as one of their feasible realizations. From this standpoint, it is reasonable to expect to find bound states for nonzero angular momentum for a catenoid covered with a hyperbolic metamaterial thin film.

\begin{figure}[htp]
    \centering
    (a)\includegraphics[width=0.4750\columnwidth]{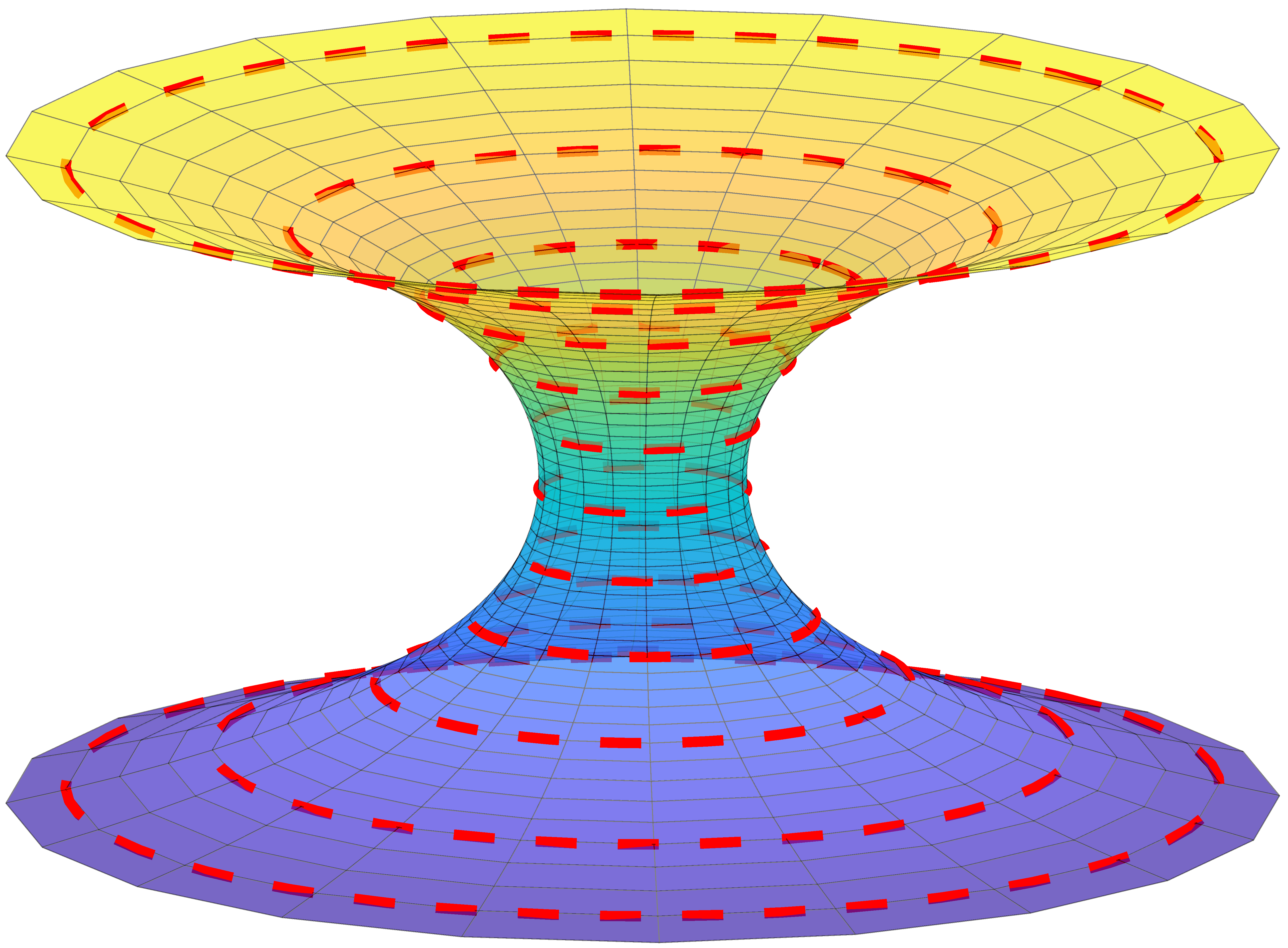}
    (b)\includegraphics[width=0.4150\columnwidth]{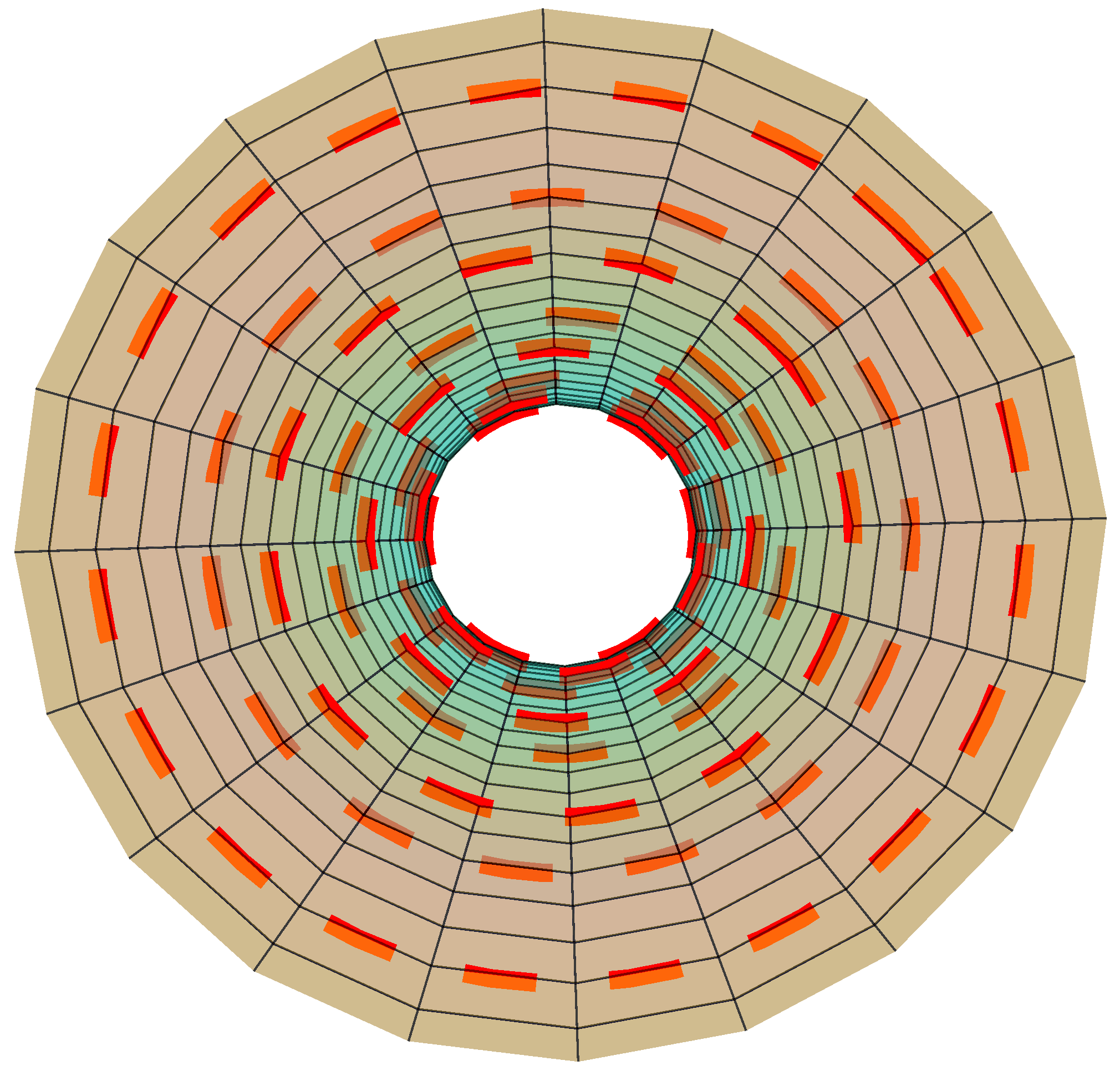} 
     \caption{(a) Hyperbolic  metamaterial  on the catenoid, with nematic liquid crystal molecules and metallic nanorods  circularly aligned as indicated by the dashed circles.  (b) Same device as as in (a) viewed from the top.}
    \label{catenoiddisclination}
\end{figure}

In this work, by using the optical metric, we study the trajectories and wave behavior of light through a thin, nematic-based, hyperbolic metamaterial film on a catenoid. We assume the director field lines oriented as shown in Fig. \ref{catenoiddisclination}. Away from the throat of the catenoid, since the surface is asymptotically flat, the director field will look like that of a disclination on a plane with a hole (similar to a $z$ constant section of the cylindrical device described in Refs. \cite{figueiredo2017cosmology,azevedo2018optical}). Here, we use a geometrical method to obtain the optical metric that describes the propagation of light along the nematic film on the catenoid, as in Ref. \cite{azevedo2021optical}. By specifying the director field $\vec{n}$ of the molecular arrangement, we can study optics in the device. From the optical point of view, this device is (locally) a uniaxial medium with ordinary and extraordinary indices given in terms of the permittivities $n_{o}^{2} = \epsilon_{\perp}$ and $n_{e}^{2} = \epsilon_{\parallel}$ \cite{kleman2007soft}. We remind that $\perp$ and $\parallel$ refer to the direction of the electric field of the light with respect to the director field. The material  is  made  of an ordinary  nematic  liquid  crystal with  an  admixture  of  metallic nanorods aligned  along  the  director field, resulting in a negative component of the permittivity ($\epsilon_{\parallel} < 0$) \cite{xiang2012liquid}. 

A coordinate system on the catenoid can be conveniently obtained from its definition in terms of the parametric equations
\begin{equation}
    \begin{split}
        x & = b_{0}\cosh(z/ b_{0})\cos\phi, \\
        y & = b_{0}\cosh(z/ b_{0})\sin\phi, \\
        z & = z, 
    \end{split}
    \label{cor}
\end{equation}
where $ b_{0}$ is the throat radius and $\phi \in \left[0, 2\pi \right]$. Intersections of $\phi = const.$ planes with the catenoid are catenaries, whose arc lengths measured from the throat ($z=0$) are given by $\tau = b_{0}\sinh(z/ b_{0})$.  This defines the coordinate system on the catenoid, composed of $\tau$ and $\phi$. 
Following Ref. \cite{azevedo2021optical}, we obtain  for the device shown in Fig. \ref{catenoiddisclination} the optical metric
\begin{equation}
    ds^{2} = - d\tau^{2} + \alpha^{2} (\tau^{2} + b_{0}^{2}) d\phi^{2},
    \label{metric1.1}
\end{equation}
where 
\begin{equation}
   \alpha = \sqrt{\frac{ \epsilon_\perp }{|\epsilon_\parallel| }}.
   \label{alpha}
\end{equation}
The metric  \eqref{metric1.1} is very similar to the metric of the device  using an ordinary nematic liquid crystal of Ref. \cite{azevedo2021optical}. But, due to the introduction of the metallic nanorods and consequently, an imaginary extraordinary refractive index, we obtain a $(-,+)$ signature of the metric instead  $(+,+)$ as before.

Now, we investigate the trajectories of light traveling on the metamaterial  film on the catenoid. For this, we start taking the geodesic equation
\begin{equation}
    \frac{d^{2}x^{\alpha}}{d\lambda^2} + {\Gamma^{\alpha}}_{\mu\nu} \, \frac{d x^{\mu}}{d\lambda} \, \frac{d x^{\nu}}{d\lambda} = 0, 
    \label{G1}
\end{equation}
where $\lambda$ is a continuous parameter, and
\begin{equation}
    {\Gamma^{\alpha}}_{\mu\nu}
  = \frac{1}{2} g^{\alpha\beta} \left(\frac{\partial g_{\beta\mu}}{\partial x^{\nu}} + \frac{\partial g_{\beta\nu}}{\partial x^{\mu}} - \frac{\partial g_{\mu\nu}}{\partial x^{\beta}} \right).
\end{equation}
From the line element given in Eq. \eqref{metric1.1}, we obtain the metric tensor as
\begin{equation}
    g_{\mu\nu} =
    \begin{pmatrix}
        -1 & 0 \\
        0 & \alpha^2(\tau^2+b_0^2)
    \end{pmatrix},
    \label{metric_tensor}
\end{equation}
such that the Christoffel symbols are 
\begin{equation}
    {\Gamma^{\tau}}_{\phi\phi} = \alpha^2 \tau
    \label{C1}
\end{equation}
and 
\begin{equation}
    {\Gamma^{\phi}}_{\tau\phi} ={\Gamma^{\phi}}_{\phi\tau} = \frac{\tau}{\tau^2 + b_0^2}.
    \label{C2}
\end{equation}
Substituting Eqs. \eqref{C1} and \eqref{C2} into Eq. (\ref{G1}), we obtain the geodesic equations
\begin{equation}
     \frac{d^{2}\tau}{d\lambda^2} + \alpha^2 \tau \, \bigg(\dfrac{d \phi}{d\lambda}\bigg)^{2} = 0
     \label{geodesic_eq1}
\end{equation}
and
\begin{equation}
     \dfrac{d^{2}\phi}{d\lambda^2} + \bigg(\frac{2\tau}{\tau^2+b_0^2}\bigg) \, \dfrac{d \tau}{d\lambda} \, \dfrac{d \phi}{d\lambda} = 0,
     \label{geodesic_eq2}
\end{equation}
where $\tau \equiv \tau(\lambda)$ and  $\phi\equiv  \phi(\lambda)$. 

\begin{figure*}[htp]
	\centering
	(a)\includegraphics[width=0.59\columnwidth]{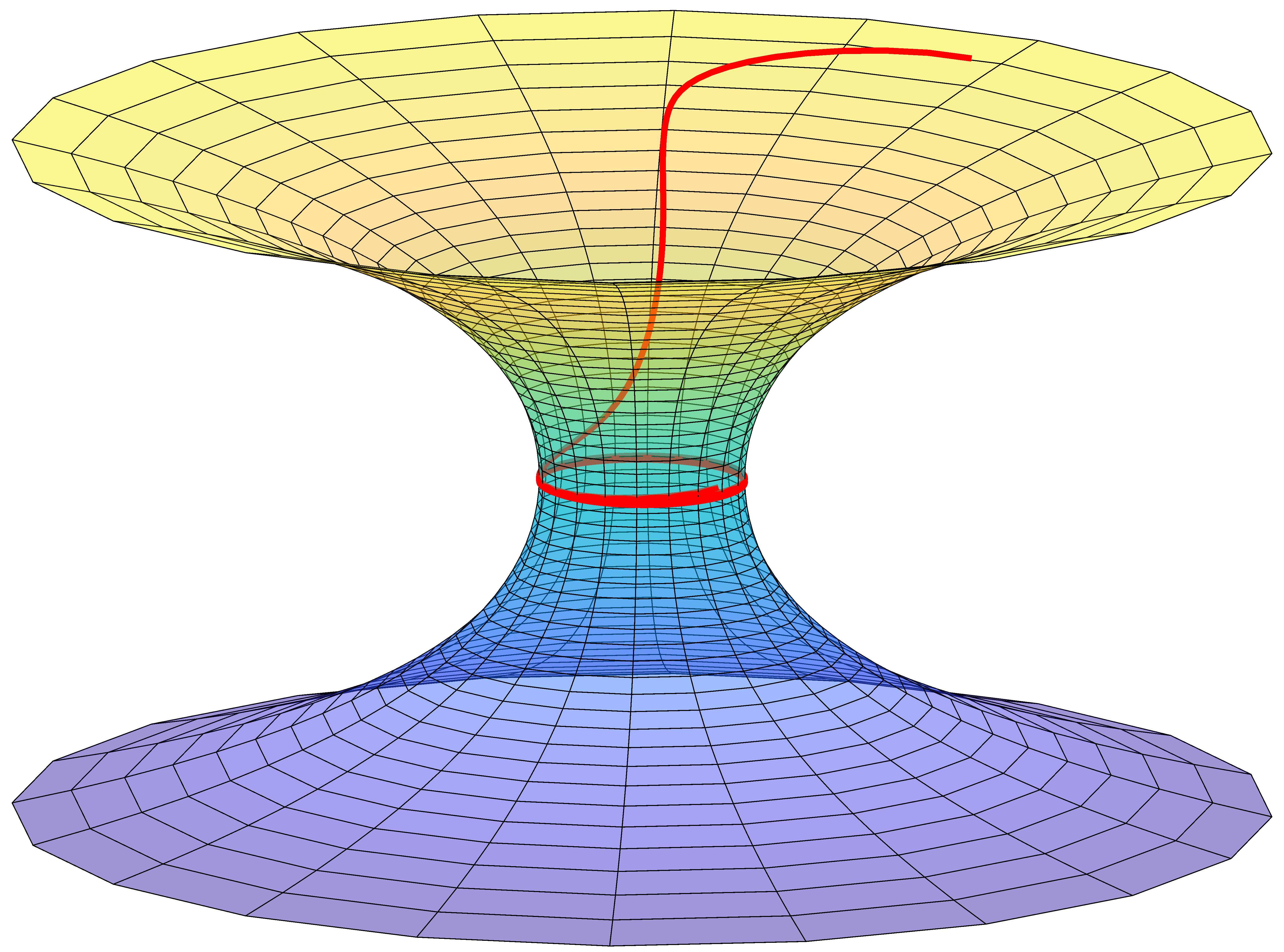}
	\hfill (b) \includegraphics[width=0.61\columnwidth]{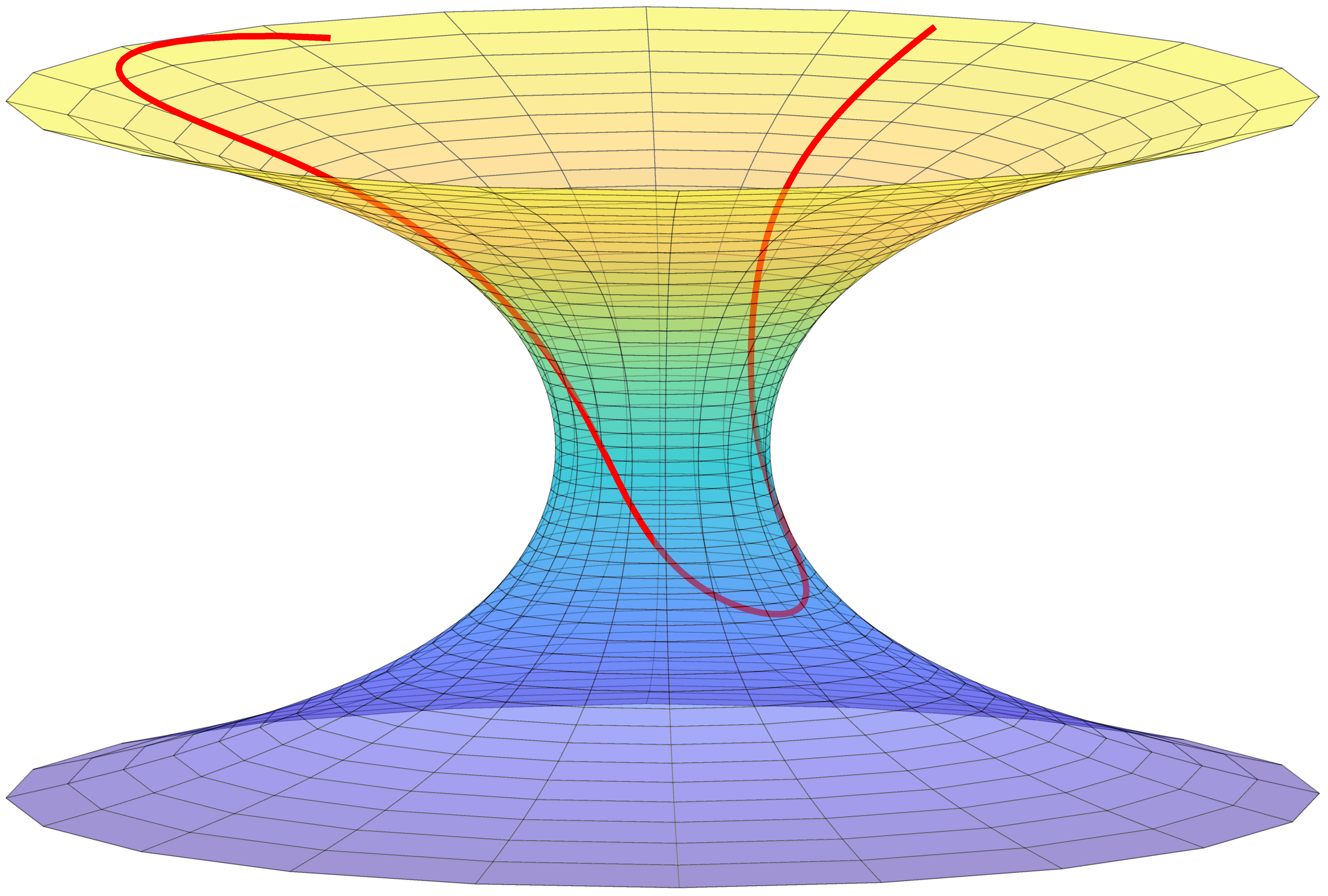}
	 \hfill (c)\includegraphics[width=0.65\columnwidth]{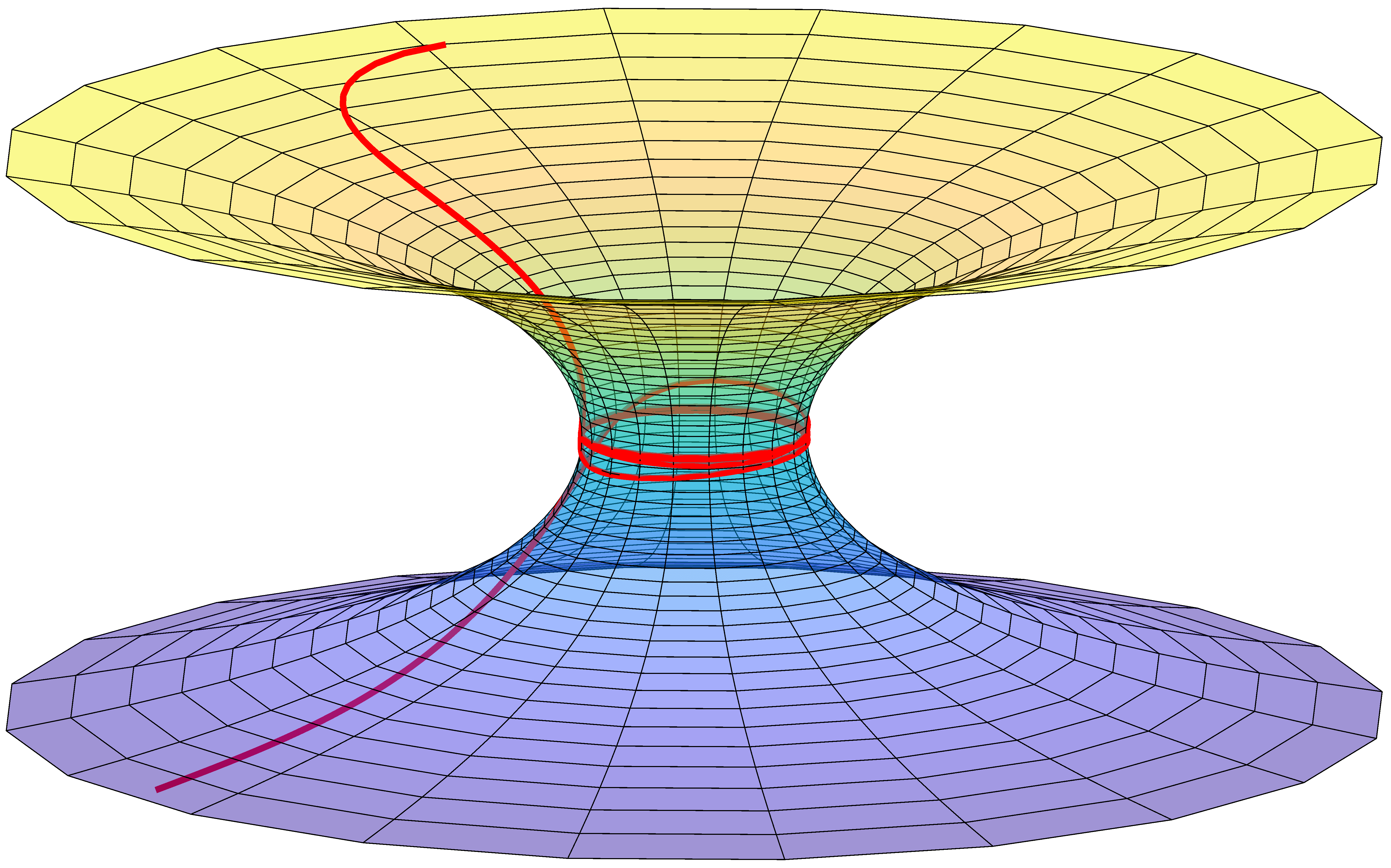}
	\caption{Trajectories of light, obtained numerically from Eqs. \eqref{geodesic_eq1} and \eqref{geodesic_eq2}, with  $\phi(0) = 0$, $\phi'(0) = 1$, $\tau(0) = 5$, and the following values of $\tau'(0)$ (which sets the shooting angle) for incident light rays: (a) $\tau'(0) = 0.1$, (b)  $\tau'(0) = 2$, and (c) $\tau'(0) = - 0.5$.}
	\label{orbs}
\end{figure*}

We numerically solve Eqs. \eqref{geodesic_eq1} and \eqref{geodesic_eq2} to find geodesics for some chosen initial conditions. The geodesics are shown in Fig. \ref{orbs}. {The numeric computational procedure employed here to solve those equations is based on Neural Networks \cite{Fotiadis1998,MICHOSKI2020193}. In particular, the python library NeuroDiffEq \cite{Giovanni2020} was used with the boundary conditions described in Fig. \ref{orbs}.} Figure \ref{orbs}(a) shows the case of light being wrapped around the throat of the catenoid. In Fig. \ref{orbs}(b), light is reflected while in Fig. \ref{orbs}(c), light winds around the throat, but it is still transmitted. The values of $b_0$ and $\alpha$ used were, respectively, 1.0 and 0.85. 

Now, to examine the  wave behavior of the extraordinary rays, we begin with the scalar Helmholtz equation for monochromatic light propagating on a curved surface \cite{schultheiss2010optics}
\begin{equation}
    \left(\Delta_{g} + k^{2}\right)\Psi = -(H^2 -K)\Psi,
    \label{waveeq}
\end{equation}
where $\Delta_{g}$ is the Laplace-Beltrami operator, $H$ is the mean curvature, $K$ is the Gaussian curvature, $\Psi$ is the amplitude of the electric field, and $k$ is the wavenumber. For the catenoid, $H=0$, and $|K| \sim 10^{-14}$ nm$^{-2}$ for surfaces of the size of centimeters (10$^{7}$ nm). Since for visible light $k^2 \sim 10^{-4}$ nm$^{-2}$, we can neglect the right-hand side of Eq. \eqref{waveeq}. Using the metric tensor \eqref{metric_tensor}, we get 
\begin{equation}
    \Delta_{g} = -\dfrac{\partial^{2}}{\partial \tau^{2}} - \dfrac{\tau}{\tau^{2} + b_{0}^{2}}\dfrac{\partial}{\partial\tau} + \dfrac{1}{\alpha^{2} (\tau^{2} + b_{0}^{2})}\dfrac{\partial^{2}}{\partial\phi^{2}},
\end{equation}
 and from Eq. \eqref{waveeq}, we find 
\begin{equation}
   - \dfrac{\partial^{2}\Psi}{\partial \tau^{2}} - \dfrac{\tau}{\tau^{2} + b_{0}^{2}}\dfrac{\partial\Psi}{\partial\tau} + \dfrac{1}{\alpha^{2} (\tau^{2} + b_{0}^{2})}\dfrac{\partial^{2}\Psi}{\partial\phi^{2}} + k^{2}\Psi = 0, \label{tacsimilar}
\end{equation}
which, with the \textit{ansatz} $\Psi(\tau,\phi) = e^{im\phi}Z(\tau)$, where $m = 0,\pm 1,\pm 2,\dots$, becomes 
\begin{equation}
    \dfrac{d^{2}Z}{d\tau^{2}} + \dfrac{\tau}{\tau^{2} + b_{0}^{2}}\dfrac{dZ}{d\tau} + \left[ \dfrac{m^{2}}{\alpha^{2}(\tau^{2} + b_{0}^{2})}-k^{2} \right]Z = 0.
    \label{tau-catenoid}
\end{equation}
The solution to this equation can be obtained in terms of the modified Mathieu function since, by substitution of $\tau = b_{0}\sinh\left({z}/{b_{0}}\right)$ in Eq. \eqref{tau-catenoid}, one obtains the modified Mathieu equation \cite{abramowitz1948handbook,azevedo2021optical}. 

\begin{figure}[htp!]
    \centering 
     (a) \includegraphics[width=0.9\columnwidth]{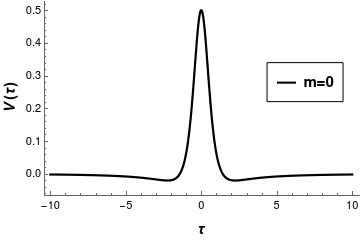} \\
     (b) \includegraphics[width=0.9\columnwidth]{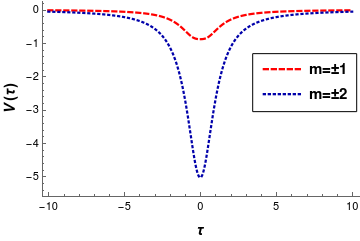} 
    \caption{Effective potential, as given by Eq. \eqref{effective}, for $m=\pm1,\pm2,\pm3$ with $\alpha = 0.85$ and $b_{0} = 1$. In graph (a), for $m = 0$, there are two minima at $\tau = \pm\sqrt{5}$  and a maximum at $\tau = 0$. In graph (b), for $m =\pm1$ and $m =\pm2$, the potential has only a minimum at $\tau = 0$. For the three curves, the potential goes to zero asymptotically.}
    \label{fig:effpot}
\end{figure}

We can write Eq. \eqref{tau-catenoid} as a Schr\"odinger-like equation \cite{kar1994scalar} by making $Z(\tau)=(\tau^{2} + b_{0}^{2})^{-1/4}\chi(\tau)$, such that 
\begin{equation}
    \dfrac{d^{2}\chi}{d\tau^{2}}-\left[k^{2} + V(\tau) \right]\chi = 0,
    \label{blueucedform}
\end{equation}
with the ``effective potential'' (see Fig. \ref{fig:effpot})
\begin{equation}
    V(\tau) = \dfrac{2b_{0}^{2}-\tau^{2} }{4(\tau^{2} + b_{0}^{2})^{2}} - \dfrac{m^{2}}{\alpha^{2}(\tau^{2} + b_{0}^{2})}.
    \label{effective}
\end{equation}
The second term in Eq. \eqref{effective} works as a ``centripetal'' potential for the angular momentum $m$. The function \eqref{effective} goes to zero when $\tau \rightarrow \pm \infty$, in agreement with the flatness of the surface in this region. For $m = 0$, it has a maximum at $\tau = 0$, and two shallow minima at $\tau = \pm \sqrt{5} b_{0}$, respectively. However, the potential has always a minimum at $\tau = 0$  for $m \neq 0$. This allows the existence of bound states or trapping of the light in the throat region. We proceed now to find some of those states by applying  a variational method to \eqref{blueucedform} and then refining the obtained estimates with the neural network method described  below Eq. \eqref{geodesic_eq2}.

Following the quantum mechanics analogy provided by Eq. \eqref{blueucedform}, we consider normalized ``wavefunctions'' obeying the boundary condition $\displaystyle{\lim_{\tau \rightarrow \pm \infty}} \chi = 0$.  We also define a bound state ``energy'' $E_{n,m}=-k^2$ corresponding to the wavefunction   $\chi_{n,m}$, where $n = 1, 2, 3, \dots$ is the quantum number that describes the order of the energy eigenvalues. The eigenvalues $E_{n,m}$ are then the expectation values of the  operator $ - \dfrac{d^{2}}{d\tau^{2}} + V$, i.e.,
\begin{equation}
    \begin{split}
        E_{n,m} = & - \int_{-\infty}^{+ \infty} \chi_{n,m}^{*}\dfrac{d^{2}\chi_{n,m}}{d\tau^{2}}d\tau \\
                  &  + \int_{-\infty}^{+ \infty} \chi_{n,m}^{*} V \chi_{n,m} d\tau.
    \end{split}
    \label{expectation-value}
\end{equation}

The shape of the effective potential suggests a Gaussian as trial function for the ground state, or $n=1$ case. Including the variational parameter $\beta_{n,m} > 0$, 
 we have for $n = 1$, 
\begin{equation}
    \chi_{1,m}(\tau) = \left( \dfrac{\beta_{1,m}}{\pi b_{0}^{2}} \right)^{1/4} \exp\left( -\dfrac{\beta_{1,m}\tau^{2}}{2 b_{0}^{2} }\right).
    \label{gauss-ground}
\end{equation}
For the first excited state, which must be orthogonal to the ground state, we propose 
\begin{equation}
    \chi_{2,m}(\tau) = \left(\dfrac{4 \beta_{2,m}^{3}}{\pi b_{0}^{6}}\right)^{1/4} \tau  \exp\left( -\dfrac{\beta_{2,m}\tau^{2}}{2 b_{0}^{2} }\right),
    \label{gauss-first}
\end{equation}
obtained from the derivative of \eqref{gauss-ground} with respect to $\tau$.

\begin{figure}[htp!]
    \centering 
     (a) \includegraphics[width=0.9\columnwidth]{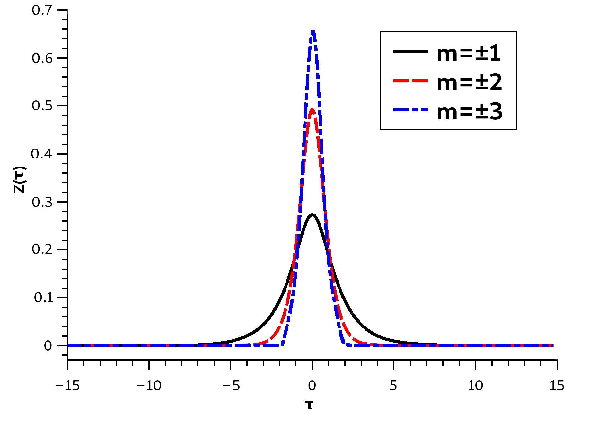} \\
     (b) \includegraphics[width=0.9\columnwidth]{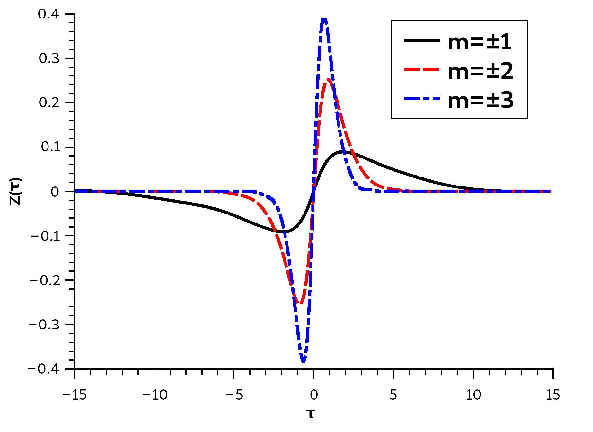} 
    \caption{(a) Radial wave functions for $n = 1$ and $m=\pm1,\pm2,\pm3$. (b) The same for $n= 2$.}
    \label{boundstates}
\end{figure}

By replacing  \eqref{gauss-ground} or \eqref{gauss-first} in Eq. \eqref{expectation-value}, we obtain the function  $E_{n,m}(\beta_{n,m})$ whose minimum provides the energy eigenvalue in each state. The aim is then to find the value of $\beta_{n,m}$ that minimizes $E_{n,m}$, for each choice of $n,m$. The wavenumbers are finally obtained from $k_{n,m} = \sqrt{- E_{n,m}}$.
 Table \ref{tab:wave-number} shows $k_{n,m}$ obtained numerically for some values of $n$ and $m$, using $\alpha = 0.85$ and $b_{0} = 1$. Using the wavenumbers supplied by Table \ref{tab:wave-number} as input, we utilized  neural networks again. This time to solve the wave equation \eqref{blueucedform}, improving then the precision of the wavefunction. Figure \ref{boundstates} shows the radial wave functions thus obtained for $n = 1$ and $n = 2$ and $m=\pm1,\pm2,\pm3$. As the wavefunctions indicate, the light rays are bound to the throat region of the catenoid. That could already be visualized in Fig. \ref{orbs}(a) obtained from ray optics.   

\begin{table}[htp!]
    \caption{Variational results for some wavenumbers $k_{n,m}$ associated with bound states of Eq. \eqref{blueucedform}.}
    \centering
    \begin{tabular}{ccc}
        \hline \hline
        $n$     &   $m$         &   $k_{n,m}$ \\ \hline
        $1$     &   $\pm 1$     &   $0.69150$ \\
        $1$     &   $\pm 2$     &   $1.85389$ \\
        $1$     &   $\pm 3$     &   $3.02896$ \\ \hline
        $2$     &   $\pm 1$     &   $0.16927$ \\
        $2$     &   $\pm 2$     &   $1.08096$ \\
        $2$     &   $\pm 3$     &   $2.17509$ \\ \hline \hline
    \end{tabular}
    \label{tab:wave-number}
\end{table}

To conclude, we want to emphasize that  we have studied here a device made from a hyperbolic metamaterial based on nematic liquid crystals under a particular arrangement.  This may lead to bound states, or trapping of light  around the throat of the device, as shown both  by ray and wave optics (respectively, classical trajectories and  the wavefunctions in the quantum mechanics analogy). This could be the starting point for the design of light storage devices. Furthermore, the trajectories that are not trapped, like the ones that are reflected and transmitted, corresponding to scattering states of the wave equation, may also have a technological application in the control of the transmission of light.    We also mention that, in Ref. \cite{azevedo2021optical}, we proposed a similar  device made from a nematic liquid crystal without the metallic nanorods which are included here. From the effective geometry given by the optical metric,  we observed there that we had an analogue  wormhole in an asymptotically flat Lorentzian (i.e. with metric signature (-1,1,1,1)) spacetime. Here, with the nanorods giving the hyperbolic metamaterial character to the system, the optical metric suggests again a wormhole but, this time in an asymptotically flat Kleinian (i.e. with metric signature (-1,-1,1,1)) spacetime \cite{figueiredo2016modeling}. Kleinian geometry leads to an alternate Special Relativity  \cite{Alves} and the possibility of using optics to simulate exotic objects in such context is an exciting research line which we are presently pursuing.

\acknowledgements
This work was partially supported by Conselho Nacional de Desenvolvimento Cient\'{i}fico e Tecnol\'{o}gico (F.M. and T.A.E.F.) and  Funda\c{c}\~{a}o de Amparo \`{a} Ci\^{e}ncia e Tecnologia do Estado de Pernambuco (J.D.M.L.).

\end{document}